\documentclass[preprint,12pt,3p]{elsarticle}

\usepackage{lineno,hyperref}
\modulolinenumbers[5]

\journal{Journal of \LaTeX\ Templates}

\usepackage{color}

 \usepackage{graphics}
 \usepackage{graphicx}
\DeclareGraphicsExtensions{.pdf,.eps,.png,.jpg,.mps}
\usepackage{amssymb}
\usepackage{booktabs}

\usepackage{multirow}

\usepackage{upgreek}

\def\Vstress{V_{\rm stress}}
\def\mVstress{\left\vert\Vstress\right\vert}

\def\td{t_{\rm stress}}

\def\DVFB{\Delta V}
\def\DQt{\Delta Q_{\rm t}}

\def\Istress{I_{\rm stress}}

\def\Ebd{E_{\rm bd}}
\def\tbd{t_{\rm bd}}

\def\Ci{C_{\rm i}}
\def\Cs{C_{\rm s}}

\def\Vac{V_{\rm ac}}
\def\Vg{V_{\rm g}}

\def\Qm{Q_{\rm m}}
\def\Qb{Q_{\rm b}}
\def\Qi{Q_{\rm i}}

\begin{document}

\begin{frontmatter}

\title{Gold/Parylene-C/Pentacene Capacitor under Constant-Voltage Stress}

\author[label1,label2]{Ibrahim~H.~Khawaji}
\address[label1]{ Department of Electrical Engineering, Pennsylvania State University, University Park, PA 16802, USA.}
\address[label2]{ Department of Electrical Engineering, Taibah University, P.O. Box 344, Al-Madina Al Munawara, KSA.}

\author[label1]{ Alyssa N. Brigeman}

\author[label3,label4]{ Osama~O.~Awadelkarim\corref{cor1}}
\cortext[cor1]{ corresponding author, ooaesm@engr.psu.edu}
\address[label3]{Department of Engineering Science and Mechanics, Pennsylvania State University, University Park, PA 16802, USA.}
\address[label4]{The Center for Nanotechnology Education and Utilization, Pennsylvania State University, 
University Park, PA 16802, USA.}

\author[label3]{ Akhlesh~Lakhtakia}

\begin{abstract}
\footnotesize
Degradation of metal-insulator-semiconductor~(MIS) capacitors of gold/Parylene-C/Pentacene 
under constant voltage stress (CVS) was investigated to explore the electrical stability and reliability of Parylene C  as a 
gate dielectric in flexible electronics. A stress voltage  of  fixed magnitude  as high as 20~V, 
both negative and positive in polarity, was applied to each MIS capacitor  
 at room temperature for a fixed duration as long as 10~s. The CVS effects on the capacitance-voltage curve-shift, the time-dependent leakage current, and the time-dependent dielectric breakdown  were measured and analyzed.  
CVS is observed to induce charge in Parylene-C and its interfaces with gold and Pentacene. The net induced charge is positive and negative for, respectively, negative and positive gate bias polarity during CVS. The magnitude of the charge accumulated following positive gate CVS is significantly higher than that following negative gate bias CVS in the range of 4 to 25 nC cm$^{-2}$. In contrast, the leakage current during the negative gate stress is three orders of magnitude higher than that during the positive gate stress for the same bias stress magnitude. The charge buildup and leakage current are due to the trapping of electrons and holes  near the Parylene-C/Pentacene interface as well as in the Parylene-C layer.~Before the application of the CVS, a dielectric breakdown occurs at an electric field of 1.62~MV~cm$^{-1}$. After the application of the CVS, the breakdown voltage decreases and  the density of the trapped charges increases  as the stress voltage increases in magnitude,   with the  polarity of 
the trapped charges   opposite to that of the stress voltage.
The magnitude and direction of the capacitance-voltage curve-shift depend on the trapping and recombination of electrons and holes  in the Parylene-C layer and in the
proximity of the Parylene-C/Pentacene interface
  during CVS.

\end{abstract}

\begin{keyword}
\footnotesize
Parylene~C, flexible electronics,   dielectric polymer, constant-voltage stress, charge buildup.

\end{keyword}

\end{frontmatter}


\section{Introduction}
\label{sec1}

The stability of flexible devices is a  major reliability concern. To enhance environmental stability,  passivation/encapsulating layers are used. Another critical reliability concern is device stability against electrical stress. Flexible devices  often operate under continuous voltage biases which can affect performance dramatically~\cite{Marszalek2017,Lee2014}. The general reason for bias instability is charge trapping~\cite{Lee2014}; electrode, insulator, and semiconductor  interfaces  easily trap charges, resulting in device-parameter shifts and degraded performances. Therefore, processes that take place at bimaterial interfaces must be characterized and controlled. 

Parylene~C is commonly used as a dielectric material for electronic applications~\cite{Marszalek2017,KhawajiTED2017,KhawajiFPE2017, Khawaji2019, Chindam2015,Chindam2014,Hsu2009,Shin2017}.~Pinhole-free films of this polymer function well as protection layers\cite{Xie2014}~to minimize device degradation caused by exposure to air and moisture. In recent years, this polymer has been used 
for flexible substrates due to its desirable mechanical properties (yield strength of 55.1 MPa and static Young's modulus of 2.76 GPa) and the ease of deposition in both micrometer- and nanometer-thickness regimes~\cite{SCS}. In addition, we have recently shown that Parylene-C columnar microfibrous thin films are viable candidates for use as interlayer dielectrics~\cite{KhawajiTED2017,KhawajiFPE2017, Khawaji2019}. 

Moreover, Parylene-C thin films exhibit superior electrical insulation characteristics, high breakdown strength ($\sim$2.5 MVcm$^{-1}$), and low dielectric loss~\cite{SCS}, not only as passivation layers but also as gate dielectrics for organic field-effect transistors (OFETs) \cite{Jakabovic2009,Tewari2009,Park2008}.  In fact, there is a limited number of materials that can be simultaneously used as a substrate, gate dielectric, and encapsulation layer simultaneously while exhibiting a high-performance level comparable to  materials dedicated to a specific application~\cite{Marszalek2017,Lee2014}.

Parylene-C films have been extensively studied \cite{Marszalek2017, Jakabovic2009,Tewari2009,Park2008, Sheraw1999, Podzorov2003, Yasuda2003, Hulea2006,  Kawasaki2010, Fukuda2012} as gate dielectrics in OFETs. However, very little is known about the electrical stability and reliability of Parylene C and its interfaces with the active layers in these devices. Current-voltage measurements are commonly made to study the electrical degradation of OFETs under constant-voltage stress (CVS) \cite{Fukuda2012}.~However, capacitance-voltage measurements are more sensitive than current-voltage measurements for investigating interface characteristics \cite{Nigam2013}. In order to address this shortfall, we investigated the electrical stability and reliability of Parylene C  as a 
gate dielectric using capacitance-voltage measurements.

	We undertook a systematic analysis of gold/Parylene-C/Pentacene metal--insulator--semi\-conductor (MIS) capacitors.~In this paper, the effects of CVS, the capacitance-voltage curve-shift, and time-dependent dielectric-breakdown (TDDB) are experimentally analyzed. Needless to add, the
gold/Parylene-C/Pentacene  system is the heart of the OFET.

\section{Experimental Procedures}
\label{sec2}	

P-type silicon ($p$-Si) substrates were ultrasonically cleaned in an ultrasonic cleaner (2200 Branson, Emerson, St. Louis, MO, USA) using, successively, acetone, DI water, isopropyl alcohol, and DI water for 10 min each. Subsequently, the cleaned substrates were etched for 20 min using a 1:4 mixture of hydrofluoric acid and DI water.

	A 150-nm-thick layer of silicon dioxide (SiO$_{2}$) was then deposited atop the  $p$-Si substrate using a PECVD tool (P-5000, Applied Materials, Santa Clara, CA, USA). Thereafter, a 50-nm-thick adhesion layer of chromium (Cr) was deposited using a sputter tool (Desktop Pro\textregistered, Denton Vacuum, Moorestown, NJ, USA). Next, a 150-nm-thick gold (Au) layer was sputtered on top of the Cr layer using the same tool.

	Pentacene  was purified in gradient-temperature sublimation system (MK-5024-S, Lindberg Electric,~Watertown,~WI,~USA) and then loaded in a tubular chamber.  A thermal gradient was maintained along that chamber maintained at about 10$^{-5}$ Torr pressure.  The material was sublimated at 300~$^{\circ}$C and re-condensed down the tube at 165~$^{\circ}$C in order to drive out impurities.  A 150-nm-thick layer of purified Pentacene was then deposited atop the Au layer via vacuum thermal evaporation (Amod, Angstrom Engineering, Kitchener, ON, Canada) at a rate of 0.2~nm~s$^{-1}$.  During that process, the Au/Cr/SiO$_{2}$/$p$-Si structure was fixed to a rotating chuck and maintained at 0~$^{\circ}$C in a chamber with a base pressure of $10^{-7}$ Torr.

	A 200-nm-thick insulating layer of Parylene C was deposited on top of the Pentacene layer using a physicochemical vapor deposition technique. An aluminum-foil boat loaded with 0.2 g of commercial Parylene-C dimer (980130-C-01LBE, Specialty Coatings and Systems, Indianapolis, IN, USA) was placed inside the vaporizer of a Parylene Labcoater (PDS2010, Specialty Coatings and Systems, Indianapolis, IN, USA). Parylene-C dimer was first vaporized at 175~$^{\circ}$C and then pyrolyzed into a reactive-monomer vapor at 690~$^{\circ}$C. The reactive-monomer vapor was diffused onto the Pentacene/Au/Cr/SiO$_{2}$/$p$-Si structure attached to a rotating platform in a vacuum chamber maintained at 28 mTorr pressure. 

	Finally, a 150-nm-thick circular disk of Au was sputtered on top of the Parylene-C insulating layer using a shadow mask to form the gate of an  Au/Parylene-C/Pentacene capacitor above the Au/Cr/SiO$_{2}$/$p$-Si bottom structure. The area of the circular disk was set as 7.1$\times$10$^{-2}$ cm$^{2}$.

	Capacitance-Voltage (C-V) measurements were carried out using a Precision LCR Meter (HP 4284, Hewlett-Packard, Palo Alto, CA, USA), while the parallel mode of `C-D' option was selected. These measurements were made at 100 kHz frequency and room temperature with an applied gate voltage $\Vg\in[-2,2]$~V   and an oscillating voltage signal $\Vac = 24$~mV.

	Using a Semiconductor Parameter Analyzer (HP 4155C, Hewlett-Packard, Palo Alto, CA, USA), we measured the time-dependent leakage current $\Istress$ as a function of  time $t$ while a stress voltage $\Vstress\in\left\{\pm10,\pm15,\pm20\right\}$~V was being  applied at room temperature.   A fresh Au/Parylene-C/Pentacene/Au/Cr/SiO$_{2}$/$p$-Si structure was used for every value of $\Vstress$.

\section{ Results}
\label{sec3}
\subsection{Capacitance-voltage characterizations before CVS}
\label{subsec1}

\begin{figure}[ht!]
\begin{center}
\includegraphics[width=3.4in]{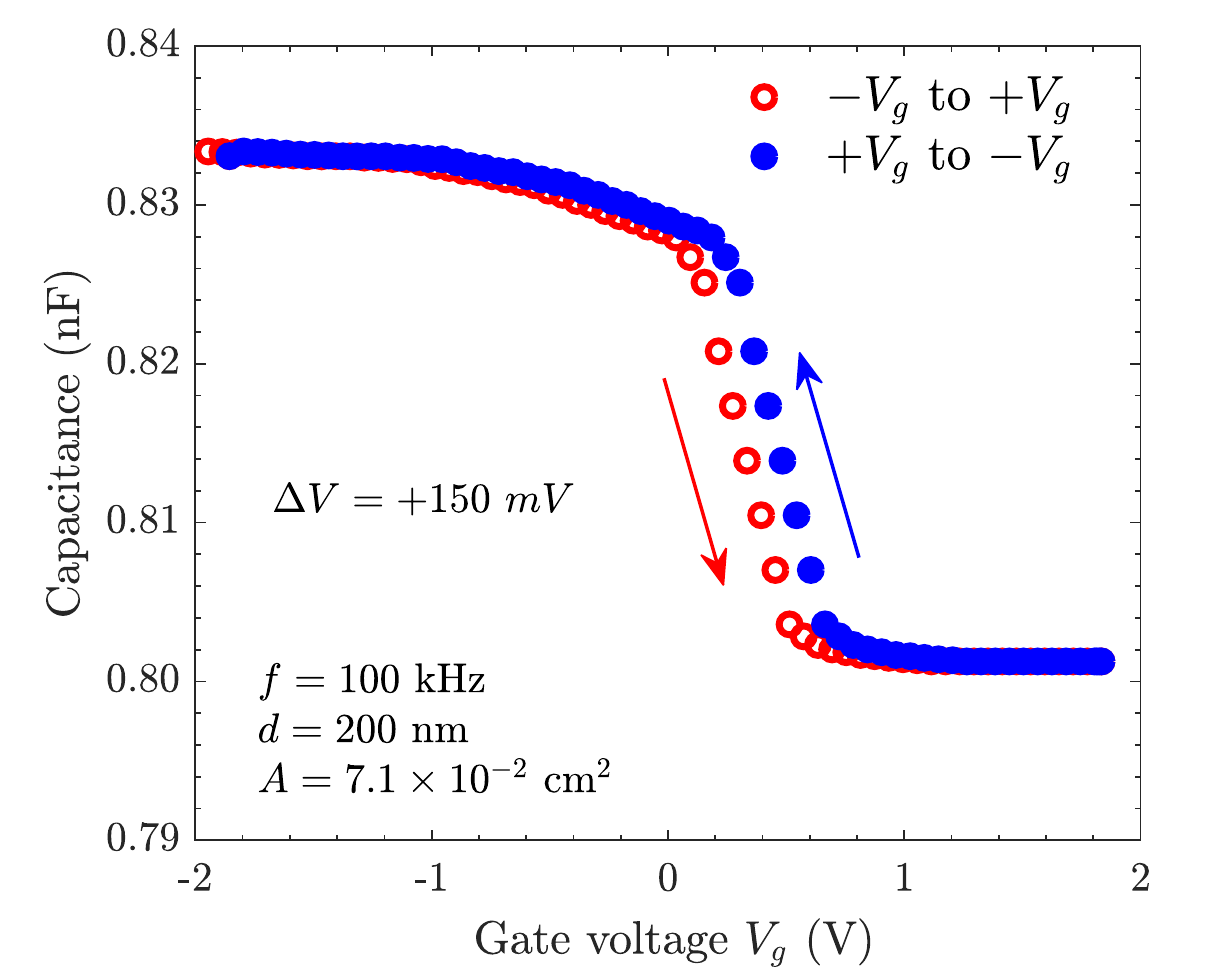}
\caption{Measured C-V characteristics of a Au/Parylene-C/Pentacene   capacitor,
functioning as a control sample (i.e., $\Vstress=0$) at 100 kHz and room temperature. The  Parylene-C layer is of thickness $d= 200$~nm and the top-electrode area
$A= 7.1\times 10^{-2}$~cm$^{2}$.}
\label{fig1}
\end{center}
\end{figure}

The measured C-V characteristics of an Au/Parylene-C/Pentacene  capacitor at 100 kHz are shown in Fig.~\ref{fig1}. As this sample was not subjected to CVS, it served as our control sample. 
Its C-V characteristics were measured in  small voltage sweeps from $\pm2$~V to $\mp2$~V.  
The capacitance shows an apparent transition from accumulation to depletion. A small hysteresis is evident, the C-V curve-shift  $\DVFB$ of 150~mV being very small \cite{Neamen2012,Hu2010}.


An MIS capacitor is generally modeled as two capacitors in series: the insulator capacitance $\Ci$ and the semiconductor-depletion-layer capacitance $\Cs$ \cite{Nigam2013, Neamen2012, Wondmagegn2011}; hence, the capacitance
\begin{equation}
 {C}= \frac{\Ci\Cs}{\Ci+\Cs}\,.
 \label{eq1-AL}
\end{equation}
For $V_g > 0$, the measured capacitance reaches a constant value equivalent to $C$ given by Eq. \ref{eq1-AL}. In contrast,  for $V_g < 0$, the capacitance saturates to a value close to 
\begin{equation}
\Ci= \epsilon_{o}\kappa \frac{A}{d}\,,
 \label{eq2n-AL}
\end{equation}
where $\epsilon_{o}=8.854\times10^{-14}$~F~cm$^{-1}$ is the permittivity of vacuum.
The Parylene-C layer is of thickness $d = 200$ nm and the top-electrode area $A=7.1\times 10^{-2}$~cm$^{2}$.


\begin{figure}[ht!]
\begin{center}
\includegraphics[width=3.4in]{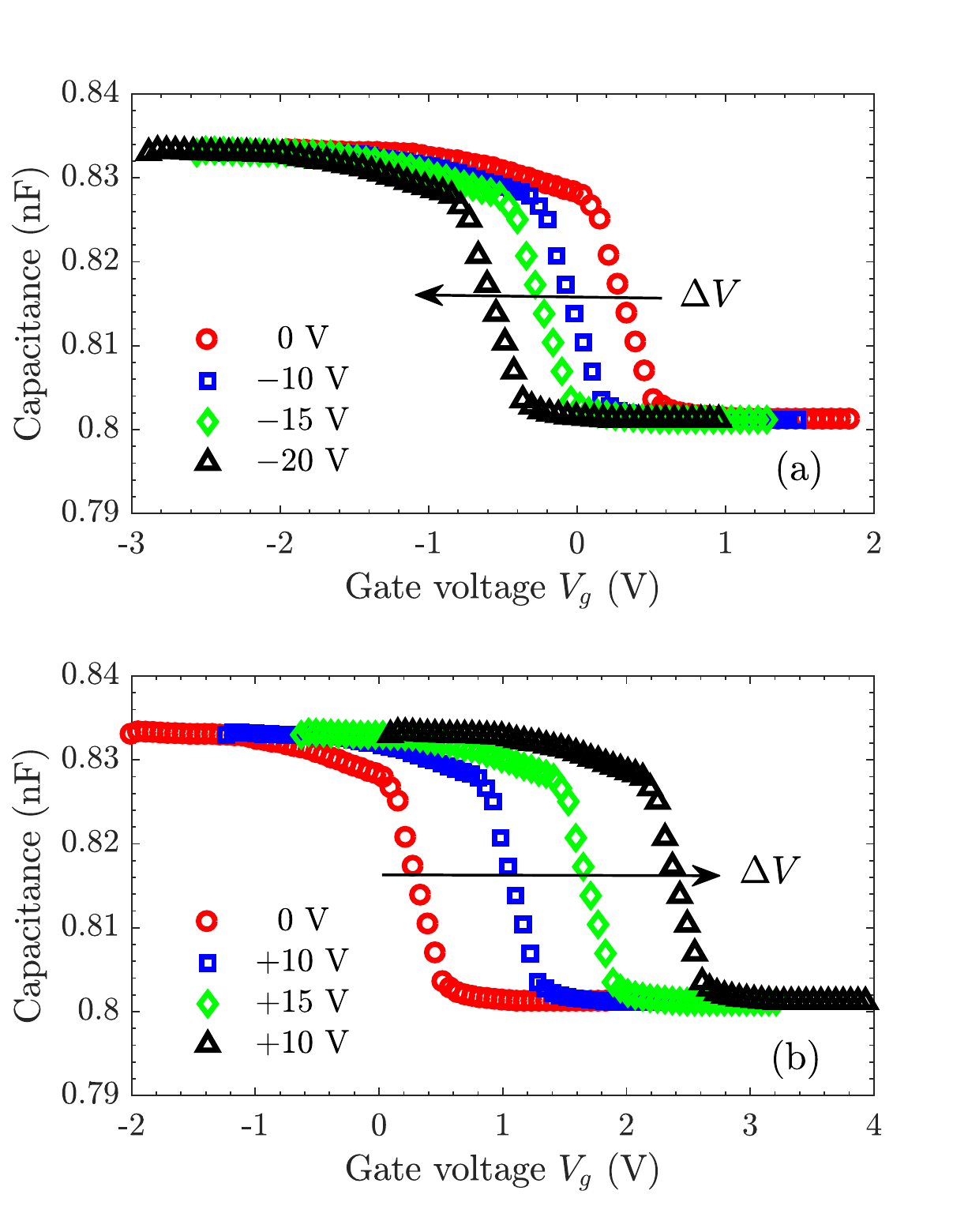}
\caption{\label{fig2} 
Measured C-V characteristics of a Au/Parylene-C/Pentacene   capacitor at 100 kHz and room temperature, after the application of (a) $ \Vstress\in \left \{ -10,-15,-20\right \}$~V and (b) $ \Vstress\in \left \{ 10,15,20\right \}$~V for $\td=10$~s.
}
\end{center}
\end{figure}


\begin{figure}[ht!]
\begin{center}
\includegraphics[width=3.4in]{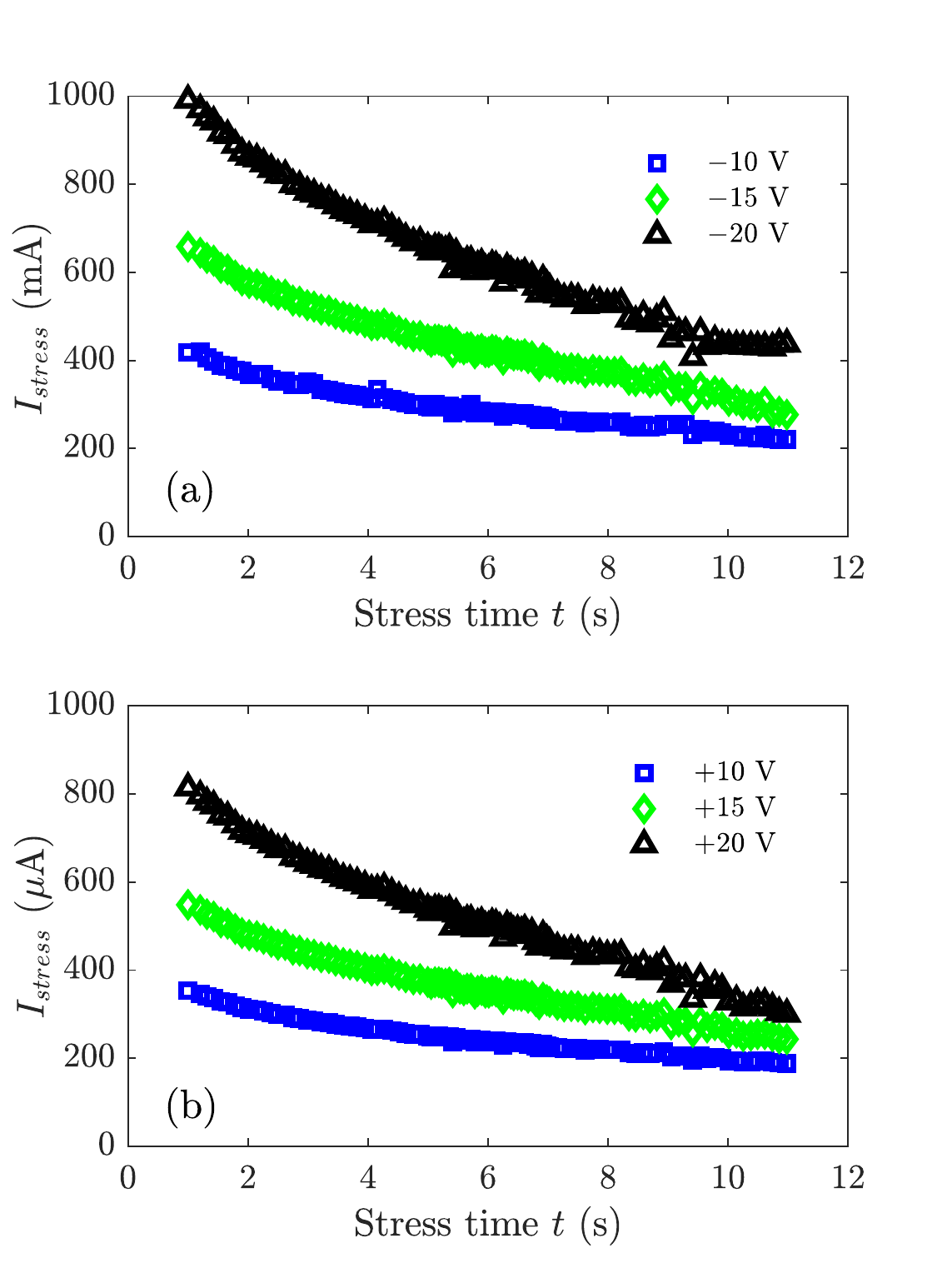} 
\caption {Measured time-dependent leakage current $\Istress$  in a Au/Parylene-C/Pentacene capacitor at room temperature as a function of time $t\in[1,11]$~s during CVS application. (a) $ \Vstress\in \left \{ -10,-15,-20\right \}$~V and (b) $ \Vstress\in \left \{ 10,15,20\right \}$~V.}
\label{fig3}
\end{center}
\end{figure}

\subsection{MIS capacitor characterizations under CVS}
\label{subsec2}
\subsubsection{Capacitance-voltage characterizations after CVS}
\label{subsec21}

Figure~\ref{fig2} shows the C-V curves measured at 100 kHz in Au/Parylene-C/Pentacene  capacitors after the room-temperature application of (a) $ \Vstress\in \left \{ -10,-15,-20\right \}$~V and (b) $ \Vstress\in \left \{ 10,15,20\right \}$~V for  duration $\td=10$~s. After the 10-s application of $ \Vstress\gtrless0$,
a C-V curve-shift $\DVFB\gtrless0$ is observed for all three values of $\mVstress$, which suggests a positive (resp. negative) charge buildup in the insulator of the MIS capacitor during the 
application of positive
(resp. negative) CVS. For the same $\mVstress$ and $\td$, $\vert\DVFB\vert$ is higher for $ \Vstress>0$ than
for $\Vstress<0$.

\subsubsection{Time-dependent leakage current under CVS}
\label{subsec22}

The time-dependent leakage current $\Istress$ was measured as a function of 
$t\in[1,11]$~s during 
the application of CVS on  MIS  capacitors   for $\Vstress\in \left \{ \pm10,\pm15,\pm20\right \}$~V. The data presented in Fig.~\ref{fig3} show that
$\Istress$ is larger for larger $\mVstress$ . For the same $\mVstress$, $\Istress$ is lower
by three orders of magnitude for $\Vstress>0$ than for $\Vstress<0$. Irrespective of 
the polarity of $\Vstress$,  $\Istress$  decays as $t$ increases.

\subsubsection{TDDB characterization after CVS}
\label{subsec23}

\begin{figure}[ht!]
\begin{center}
\includegraphics[width=3.4in]{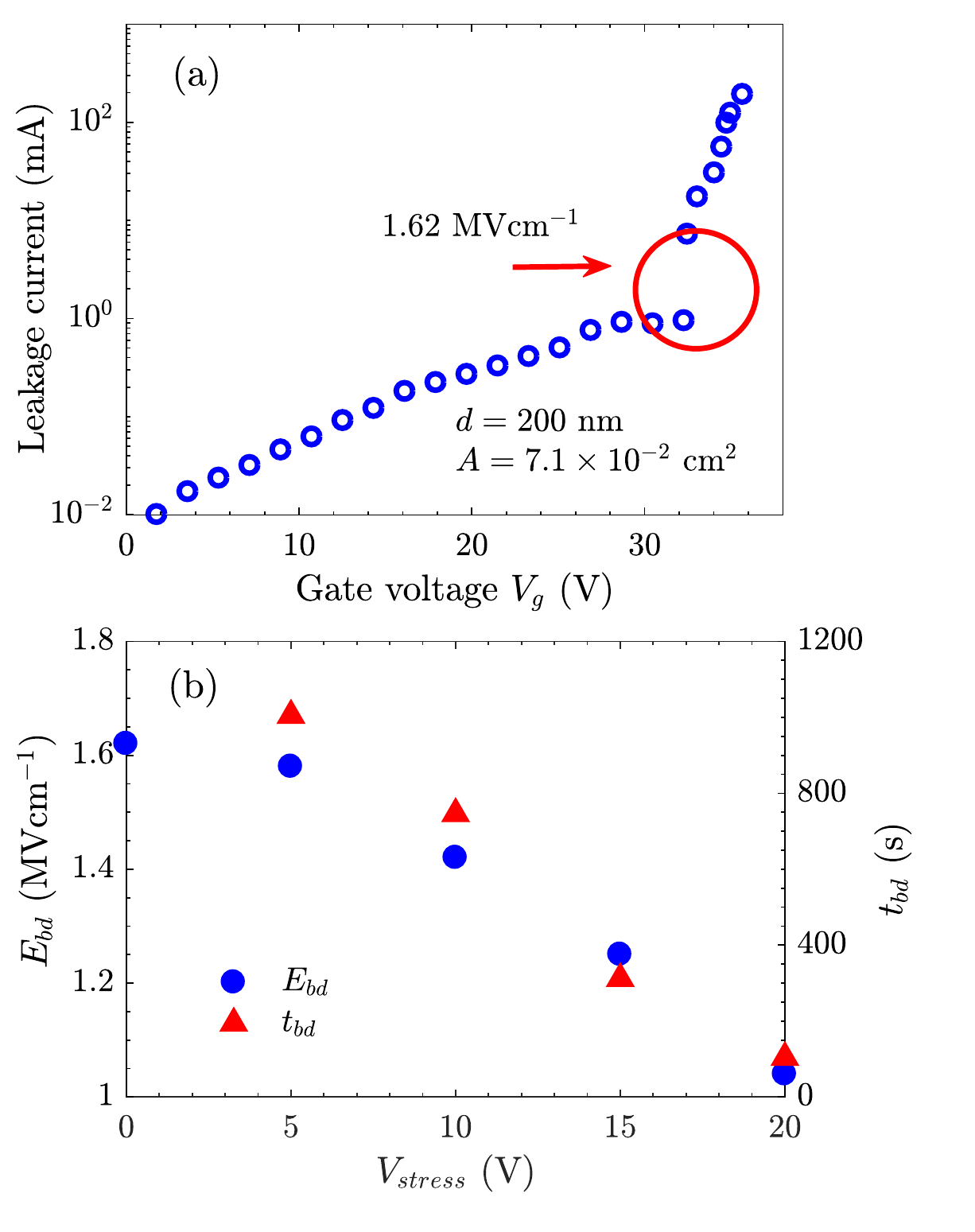}
\caption{Room-temperature TDDB characteristics  of
Au/Parylene-C/Pentacene  capacitors: (a)  Leakage current measured as
a function of gate voltage $\Vg$ before CVS application (control sample).
(b) Breakdown electric field $\Ebd$ and time $\tbd$ after application of $ \Vstress\in \left \{5,10,15,20\right \}$~V. }
\label{fig4}
\end{center}
\end{figure}

The TDDB characteristics of the Au/Parylene-C/Pentacene  capacitors were obtained by recording the current-voltage (I-V) response before and after the 10-s application of $ \Vstress>0$.~Before CVS, the 
I-V curve  in Fig.~\ref{fig4}(a) shows  a dielectric breakdown occurs at an electric field of 1.62~MV~cm$^{-1}$.~This value is comparable to  the values of the breakdown electric field $\Ebd$  in the range 1.9 to 2.2 MV cm$^{-1}$~reported for 200-nm-thick Parylene-C layers in MIM structures \cite{Shin2017,Gowrisanker2009}, but for a much smaller gate electrode area of about 3$\times$10$^{-4}$ cm$^{2}$.


In Fig.~\ref{fig4}(b), 
measured values of the breakdown field $\Ebd$  and  the time-to-breakdown $\tbd$ are plotted as functions of  $ \Vstress \in \left \{ 5,10,15,20\right \}$~V. Time-to-breakdown $\tbd$ is the time it takes for the Parylene C to breakdown under the application of each $ \Vstress$.

As expected, both  $\Ebd$  and $\tbd$ decrease as $ \Vstress>0$ increases. The value of $\Ebd$   decreases from 1.62 MV~cm$^{-1}$ for the control sample (i.e.,
$\Vstress=0$) to 1.04 MV~cm$^{-1}$ for the sample stressed with $ \Vstress=20$~V. Furthermore, $\tbd$ decreases from 1005~s for the control sample  to 104~s for the sample stressed with $ \Vstress=20$~V.

\section{ Discussion}
\label{sec4}

The C-V characteristics of organic-based MIS capacitors are limited by contact injection \cite{Nigam2013, Nigam2014, Diao2007, Horowitz2004, Klauk2010}.~The C-V curve obtained in Fig. \ref{fig1} can be explained in terms of  charge accumulation arising from  injection and contained within the semiconductor layer.

For $V_g > 0$, a thin accumulated layer of injected holes occurs at the  Au/Pentacene
interface, while the Parylene-C/Pentacene interface is depleted and devoid of any significant free charge carriers.~As a result, a depletion layer is created inside the Pentacene.~Hence, the capacitance is given by $C$ in  Eq. \ref{eq1-AL}.  For $V_g < 0$,
an accumulation of holes occurs near the interface of Pentacene/Parylene-C. As $V_g$ further increases to more negative values, $\Cs$  increases and $C$ approaches $\Ci$   given by Eq. \ref{eq2n-AL}.

The C-V curve shifts observed in Fig.~\ref{fig2}~are attributed to charge buildup $\DQt$ in Parylene C \cite{Schroder2006,Pumar2017}.
 Charges of three different provenances are associated with $\DQt$ \cite{Schroder2006,Pumar2017}; i.e.,
\begin{equation}
\DQt=\Qm+\Qb+\Qi\,.
\end{equation}
Here,
$\Qm$ is the charge density  of mobile positive charges  located in the bulk of the insulator and arising from ionic impurities such as {Na}$^{+}$. The effect of these charges can be seen as a hysteresis in the C-V curve when sweeping $\Vg$ in a negative-positive-negative loop.
Also, $\Qb$ is the charge density of charges trapped in the bulk insulator. It can be either negative or positive, depending on whether holes or electrons are trapped.
$\Qi$ is the charge density of charges trapped in the semiconductor/insulator interface. 
  It can also be either negative or positive. Because these charges are trapped at the interface,
  $\Qi$ has the largest effect on  $\DVFB$.

As can be deduced from Fig.~\ref{fig1}, $\Qm$ is negligibly small because a very small hysteresis ($\DVFB=150$~mV) is detected as the gate voltage is swept from $-2$~V to $+2$~V to $-2$~V.
Hence, $\Qm$ plays no role in the charge buildup observed in Parylene C after the
application of CVS so that 
\begin{equation}
\DQt=\Qb+\Qi\,.
\end{equation}

Let us now attempt a quantitative analysis of $\DQt$ and its dependence on $\Vstress$ and $\td$. Accordingly \cite{Farmer2007},

%
\begin{eqnarray}
\DQt =-\frac{\DVFB\,\Ci}{A}\,.
\label{eq5}
\end{eqnarray}
%
Table~\ref{table1}~provides the values of $\DQt$ for $\Vstress\in\left\{\pm10,\pm15,\pm20\right\}$~V.
Clearly, the charge buildup are positive for $\Vstress<0$ but negative for
$\Vstress>0$. Furthermore, for fixed $\mVstress$, the charge buildup is more than twice in magnitude for $\Vstress>0$
than for $\Vstress < 0$.


\begin{table*}[ht!]
\caption{C-V curve-shift $\DVFB$ of and the charge buildup  $\DQt$  in a Au/Parylene-C/Pentacene MIS capacitor subjected to $ \Vstress \in \left \{ \pm10,\pm15,\pm20\right \}$~V for $\td=10$~s.}
\label{table1}

\scriptsize
\centering

\begin{tabular}{cccccccc}
\toprule
     & \multicolumn{6}{c}{$\Vstress$~(V)} \\
\\ 
     & $-10$    & $-15$   &$ -20$&   & +10    & +15   & +20\\
 \hline
\\
$\DVFB$~(V) & $-0.35$      & $-0.56$    & $-0.89$&     & +0.76      & +1.37     & +2.10\\
\\
$\DQt$~(C cm$^{-2})$  & +4.10$\times10^{-9}$     & +6.50$\times10^{-9}$    & +1.04$\times10^{-8}$ &   & $-8.90\times10^{-9}$     & $-1.60\times10^{-8}$    & $-2.45\times10^{-8}$ \\

\bottomrule
\end{tabular}

\end{table*}


During the time that $\Vstress<0$, electrons are injected from the gate and holes are injected from the layer of accumulated holes near the Parylene-C/Pentacene interface. Electrons and holes transiting the Parylene-C layer can be trapped at defect sites and give rise to charge buildup. It is apparent from  Table.~\ref{table1} that hole trapping dominates and the net charge buildup is positive for $\Vstress<0$.

In contrast and during the time that $\Vstress>0$, holes are only injected from the gate into Parylene-C. The resulting hole-leakage current is observed to be much smaller than $\Istress$ shown in Fig.~\ref{fig3}(a).~This is because a significant portion of the applied $\Vstress$ is dropped across the depleted (i.e., devoid of charge carriers) layer near the Pentacene/Parylene C interface; much less hole transport takes place across Pentacene for $\Vstress>0$. However, the charge-buildup sign is negative, which may indicate that the observed charge buildup is not entirely due to the trapping of charge carriers (electrons and holes)   but could also be caused by defect generation. Presumably, these generated defects are electrons traps that are populated during the application of $\Vstress$.

The I-V curve in Fig.~\ref{fig4}(a) shows  a dielectric breakdown occurs at an electric field of 1.62~MV~cm$^{-1}$. This   breakdown may indicate the formation of a defect-related conduction path \cite{Raghavan2014, Sune2000}. In other words, a higher applied voltage could induce defects that eventually form different conducting paths from the gate to the semiconductor in the Au/Parylene-C/Pentacene  structure.

As shown in Fig.~\ref{fig4}(b), the decrease in $\Ebd$ suggests that more/longer 
 conductive paths are formed with increasing $\Vstress>0$.~Conductive paths result from defect-generation processes and, hence, more defects are presumably generated as $\Vstress$ further increases to   higher values.~This deduction is in agreement with earlier inference from Table~\ref{table1} that  defect generation dominates over charge trapping during $\Vstress>0$.

\section{Concluding Remarks}
\label{sec5}

A systematic analysis of the reliability of Au/Parylene-C/Pentacene MIS capacitors
under constant-voltage stress was performed, with focus on the effects of CVS on the stability of Parylene~C as a gate dielectric. 200-nm-thick Parylene-C thin films  were utilized as gate-dielectric layers of Au/Parylene-C/Pentacene MIS capacitors. 
Measurements and analysis of the C-V curve-shift, the time-dependent leakage current, and the time-dependent dielectric breakdown were performed before and after application of CVS. Positive and negative stress voltages of the same magnitude were applied for 10-s duration.

A summary of our results is as follows:
\begin{itemize}
\item The MIS capacitance shows an apparent transition from accumulation to depletion.
\item The C-V curve-shift  is higher for positive
 stress voltage than for negative stress voltage of the same magnitude.

\item The time-dependent leakage current for positive stress voltage is three orders of magnitude lower than for negative stress voltage.
\item The charge density of trapped charges increases with the stress voltage,
the polarity of the trapped charges being opposite to the polarity of the stress voltage.
\item As the application of CVS increases, the breakdown voltage decreases as does the time to breakdown.
\end{itemize}

Therefore, our main conclusions are as follows:
\begin{itemize}

\item The C-V characteristic can be explained in term of  accumulation charges within the semiconductor layer. 
\item This accumulation charge is due to contact injection.
\item Inside the insulating layer,the charge buildup resulting from the accumulation of  trapped charges affects the stability of the the MIS capacitor by shifting its C-V curve. 
\item The shift of the C-V curve is attributed to the  trapping and recombination of electrons and holes inside Parylene C and its interface with Pentacene.
\item The dielectric-breakdown mechanism is defect dominated. 
\end{itemize}

Overall, the buildup of trapped charges in the Parylene-C layer and near the Parylene-C/Pentacene interface plays a major role in the degradation of Au/Parylene-C/Pentacene capacitors. Our analysis   in this paper provides a first-level understanding of the charge buildup in Au/Parylene-C/Pentacene capacitors and, perhaps, will serve as the basis of future studies on the defect-generation process and the trapping of charge carriers within the insulator layer in OFETs.

\vspace{1cm}
\noindent \textbf{Acknowledgment}
~
\vspace{0.3cm}

We thank Prof.~Chris Giebink for providing access to the vacuum thermal evaporation tool in the Applied Optoelectronics \& Photonics Lab at the Pennsylvania State University.
AL is grateful to the Charles Godfrey Binder Endowment at the Pennsylvania State University for ongoing support of his research activities.

\end{document}